\documentclass[twocolumn,pre,a4paper,showpacs]{revtex4}
\usepackage{amsbsy}
\usepackage{amssymb}
\usepackage[dvips]{graphicx}

\begin{document}
LPT-07-44

\title{Network of inherent structures in spin glasses: scaling and scale-free
distributions}

\author{Z. Burda}
\affiliation{Marian Smoluchowski Institute of Physics
and Mark Kac Complex Systems Research Centre, Jagellonian University,
Reymonta 4, 30-059 Krakow, Poland}

\author{A. Krzywicki}
\affiliation{Laboratoire de Physique Th\'eorique,
b\^at. 210, Universit\'e Paris-Sud, F--91405 Orsay, France.}

\author{O. C. Martin}
\affiliation{Univ Paris-Sud, UMR8626, LPTMS, Orsay, F-91405; CNRS, Orsay,
  F-91405, France.}

\date{\today}

\begin{abstract}
The local minima (inherent structures) of a system
and their associated transition links give rise to
a network. Here we consider the topological 
and distance properties of such a network in the context
of spin glasses.
We use steepest descent dynamics, 
determining for each disorder sample the 
transition links appearing within a given 
barrier height.
We find that differences between linked inherent structures
are typically associated
with local clusters of spins; we interpret this within a framework
based on droplets in which the characteristic ``length scale''
grows with the barrier height.
We also consider the
network connectivity and the degrees of its nodes.
Interestingly, for spin glasses based on random graphs,
the degree 
distribution of the network of inherent structures exhibits a 
non-trivial scale-free tail.
\end{abstract}
\pacs{75.10.Nr, 75.40.Mg, 89.75.Fb}

\maketitle

\section{Introduction}

In typical complex systems, there is a huge number of 
inherent structures~\cite{StillingerWeber84} (local minima of 
energy in the configuration
space), with no single one giving a good approximation to the
equilibrium state. Indeed, it is necessary to consider most if
not all inherent structures and their basins to reach a quantitative
understanding of a model's equilibrium 
properties~\cite{BuchnerHeuer99,BogdanWales06}. To
include also the dynamics, it is necessary to know how these
basins are connected (linked)
by saddles, the (excitation) energies of which
play an important role for the associated transition 
rates~\cite{CrisantiRitort00,DoyeWales02,GiovambattistaStarr02,
DoyeWales03,DoliwaHeuer03,StarioloArenzon04,DennyReichman03}.
The set of local minima and their links define a
network~\cite{AngelaniParisi98,Doye02}; following Doye~\cite{Doye02}, 
we shall refer to it as the ``Inherent Structure Network'' or ``ISN''.
Such a framework
has been applied to water~\cite{GiovambattistaStarr02},
atomic clusters~\cite{Doye02} and
glasses~\cite{AngelaniParisi98};
all these systems have a somewhat complex energy landscape
but none has quenched disorder.
\par
In this work, we focus on the inherent structure network (ISN) in the 
context of spin glasses~\cite{MezardParisi87b} because they combine
quenched disorder and complex energy landscapes; in addition, their
landscapes are based on microscopic Hamiltonians rather than
on an abstract random potential~\cite{CavagnaGarrahan99,Fyodorov04}. Of 
particular interest are the network's
topology, how the linked inherent structures differ,
and the associated scaling laws with the number $N$ of spins.
The case of one-dimensional spin glasses is simplest
and we shall use it to motivate a framework based on droplets. Such a picture
predicts that the degree of the nodes in the ISN 
follows a Gaussian distribution with a mean
growing linearly with $N$.
These properties are well borne out in the one-dimensional spin glass,
but when the spins lie on a random graph, 
fat tails appear in the distribution
of the degree, indicating instead a scale-free
behavior of the ISN.
\par
The paper is organized as follows.
The models are defined in Sect.~\ref{sect_MODELS}, and we
also present our methods and the observables of interest.
In Sect.~\ref{sect_ONEDIMENSION} we investigate in 
detail the case of the one-dimensional spin glass.
Then we present a framework based on ``droplets'' and give
its predictions (Sect.~\ref{sect_DROPLET}); these are
expected to be valid when
droplets are small and ``weakly interacting''.  Finally,
in Sect.~\ref{sect_INFINITEDIMENSION} we cover
the case of spin glasses based on random graphs; there
droplets are strongly interacting and context sensitive.
We conclude in Sect.~\ref{sect_CONCLUSIONS}.

\section{Models and Methods}
\label{sect_MODELS}
\subsection{Spin-glass models}
We consider $N$ Ising spins lying on the vertices of a 
connected graph $G$ where each vertex is connected
to exactly $k$ others. To each edge $ij$ of such a graph, 
we independently assign a weight $J_{ij}$ 
according to a distribution symmetrized about 0.
These elements, i.e., the (random) edges
and their associated weights $J_{ij}$, define the
system's ``quenched disorder''. The energy landscape is defined
from the space of spin configurations via the Hamiltonian
\begin{equation}
\label{eq:H}
H(\{\sigma_{i}\}) \equiv
  -\sum_{\langle ij \rangle} J_{ij}\;\sigma_{i}\;\sigma_{j}\;,
\end{equation}
where the sum runs over all pairs of sites connected by an edge of
the graph and $\sigma_{i}=\pm 1$ is the Ising spin on site $i$.
As in ref.~\cite{BurdaKrzywicki06} 
the weights $J_{ij}$ are generated from a
Gaussian distribution with variance
\begin{equation}
\label{V}
\mbox{\rm var}(J_{ij})=\frac{J^2}{k} 
\end{equation}
where $J$ is an energy scale 
set to 1 in the numerical work; this scaling 
allows for an extensive thermodynamic limit
for all $k$, but in our case the landscape properties
do not depend on this rescaling.
The model has an obvious global $Z(2)$ 
symmetry corresponding to flipping 
all the spins simultaneously.

We shall consider two classes for $G$. The
first class is that of a one-dimensional lattice with  
nearest neighbors only and periodic boundary conditions ($k=2$);
the simplicity of this structure allows for a relatively
complete understanding of the associated ISN.
The second class corresponds to 4-regular \emph{random} graphs
where each spin is coupled to exactly four other 
ones~\cite{DominicisGoldschmidt89}; here 
a rather careful analysis is needed to compensate for the 
rather small sizes accessible to our numerical computations. 

\subsection{Inherent structures and their network}
\label{subsect_ISIN}
There are $2^N$ possible configurations of the spins. A configuration
is an {\em inherent structure} (IS) if and only if it is a 
local minimum of the energy, i.e., no single spin flip
will lower the energy. (Because the $J_{ij}$ are i.i.d.
variables with a continuous distribution, generically the energy
of an inherent structure is strictly lower than that of
its neighboring configurations.) 

Each spin configuration can be mapped onto a unique inherent 
structure using the following {\em steepest descent} procedure:
a succession of spin flips is carried out but at each step the
spin chosen for flipping is that which lowers the energy the most.
The basin of attraction (hereafter simply referred to as ``basin'') 
of an inherent structure
is, by definition, the set of all configurations mapped onto that
structure by this steepest descent procedure.

Configuration space can be thought of as forming a Boolean
hypercube: two configurations are
connected by an edge if they differ by a single spin flip. Let
$x$ and $y$ be two configurations mapped by steepest descent
onto the inherent structures
$X$ and $Y$ respectively: $x \to X$ and $y \to Y$. Suppose that $x$ 
and $y$ are connected by an edge (they 
differ by a single spin flip) and that $X \neq Y$. In 
this case we say that the $xy$ link
crosses the frontier between the basins of 
$X$ and $Y$: it is a {\em frontier link}. 

\begin{figure}
\includegraphics[width=5cm]{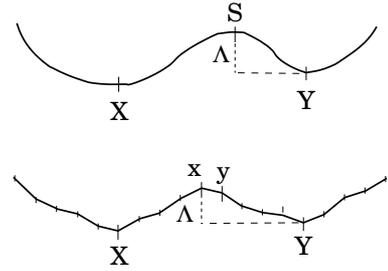}
\caption{\label{fig:link} 
Representation of an energy landscape and the associated special
configurations. Horizontal axis: configuration space,
vertical axis: energy. $X$ and $Y$ are inherent structures, connected
by a saddle $S$ (in the continuous case, top) or by a 
frontier link $xy$ (in the discrete case, bottom).}
\end{figure}

Let $xy$ be a frontier link and let $x$ be more energetic than $y$: 
$H(x) > H(y)$. The link $xy$ is a {\em transition link} if the
following condition is satisfied:
\begin{equation}
\label{eq:transitionLink}
\left[ H(x)-H(X) \right] <\Lambda  \;\; \mbox{\rm or} \;\;
\left[ H(x)-H(Y) \right] <\Lambda\; ,
\end{equation}
where $\Lambda$ is a control parameter. The situation is illustrated
in Fig.~\ref{fig:link}, both for this discrete space case
and for the analogous 
situation in a continuous space where the distinction between
$x$ and $y$ is not necessary. In this figure, 
the configuration $x$ - the 
more energetic vertex in a transition link - is the discrete space analogue
of a {\em transition state} (saddle) in models having a continuous 
configuration space (cf.~\cite{Wales03}); the saddle
in the top part of Fig.~\ref{fig:link} is denoted by $S$. There may be many 
saddles between two neighboring basins,
so one could think of having either multiple or weighted
links between inherent structures. In the present paper we are not interested
in these multiple links, but only whether there
exists at least one $xy$ between given IS which 
fulfills Eq.~(\ref{eq:transitionLink}),
in which case the two IS are connected by a link.
(In this respect we follow ref.~\cite{DoyeMassen05}).  

The ISN is defined by its nodes (the inherent structures)
and its links (two nodes are linked if the corresponding
inherent structures have basins connected by a transition link). 
It can be thought of as a directed network:
by convention the direction of an edge is from the more
energetic to the less energetic node. The topology of the ISN
depends, of course, on the choice of $\Lambda$. For $\Lambda \ll J$ the
ISN consists mostly of isolated nodes (some nodes may nevertheless be
connected due to the random values of the $J_{ij}$, which 
occasionally make certain
energy barriers small). The number of links of the ISN increases with 
$\Lambda$ and a giant component - defined disregarding the directionality
of edges - is expected to set in when $\Lambda$ is sufficiently large. 

\par
\subsection{Algorithmic methods}
\label{subsect_algorithms}
Given any realization of the graph and the weights $J_{ij}$, we exhaustively
consider all configurations and determine via the steepest descent
algorithm the basin to which each belongs. 
This gives the complete map from the $2^N$ configurations to the
set of inherent structures. 
(For computational details, see ref.\cite{BurdaKrzywicki06}.)
Then we determine the transition states which allows us to 
introduce links between inherent structures; this leads to 
the ISN. For all that follows,
two nodes of the network are connected by at most one
link, of energy given by that of the lowest
transition state between the two considered inherent structures.
To understand the statistical behavior at large $N$ of such 
ISN, we repeat this construction 
for many disorder samples and try to extract the dependence on $N$.

The CPU time needed to execute this code for one instance of 
$20, 25$ and $30$ spins is about 2s, 100s and 4000s, respectively. It is
evident that, in practice, $N \approx 30$ turns out to be a working
limit for our programs.
Moreover, the RAM memory demand is considerable: 
at $N=30$ one needs 2Gb.

\section{The one-dimensional case}
\label{sect_ONEDIMENSION}
\subsection{Inherent structures}

On the one-dimensional lattice, a configuration can be
described either by the values of every spin or by
the ``bond'' quantities $\sigma_i J_{i,i+1} \sigma_{i+1}$; these
are positive for satisfied bonds and negative for
unsatisfied ones. In this last representation, it is easy
to give a necessary and sufficient condition for having a 
local minimum of the energy: each unsatisfied bond must be between
two satisfied bonds, and its strength $|J_{i,i+1}|$
must be smaller than that of its neighbors.

\begin{figure}
\includegraphics[width=5cm]{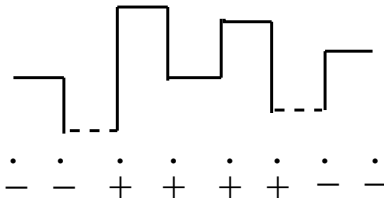}
\caption{\label{fig:landscape} 
Energy profile
representation of configurations for 
the one-dimensional lattice. The dots stand for the 
lattice vertices, the strength of a bond is given by its 
``height'' on the vertical axis and the unsatisfied bonds are 
represented by broken segments. A particular spin configuration
is indicated by the $\pm$ signs; for this 
configuration the second and the sixth
bond are unsatisfied.}
\end{figure}

Without loss of generality (at least if one neglects the periodic boundary
conditions of the lattice), one can replace all the $J_{i,i+1}$
by $|J_{i,i+1}|$. (This follows from the gauge invariance
of the Hamiltonian.) We are then led to an
``energy profile''
representation where each bond has a height $|J_{i,i+1}|$
and a configuration is specified by specifying which bonds
are unsatisfied. In Fig.~\ref{fig:landscape}
we show such a representation where unsatisfied bonds are 
shown as broken segments, and the ``height'' of the bond
is given on the vertical axis. Note that in ferromagnets,
such unsatisfied bonds are referred to as kinks.
It is straightforward
to see that a configuration is a local minimum if and only
if its broken segments (if any) appear 
only at the bottom of the 
profile. We shall call ``elementary
droplet'' the region of spins contained between two
successive minimum segments; a spin
on a minimum segment belongs to the single elementary droplet it
touches. It is easy to see that all inherent structures
correspond to simply specifying the state of each 
elementary droplet; within the gauge where all the
$J_{ij}$ are positive, each elementary droplet will have all of its
spins in the $+1$ state or all in the $-1$ state.

For a given sample, let $M$ be the number of segments
that are locally minimum in the 
energy profile. (For periodic boundary
conditions, this is also the number of elementary droplets.)
Then the number
of inherent structures is $2^M$, a well known 
result \cite{Li81,DerridaGardner86c}.
At large $N$, $M$ is a Gaussian random variable of mean
$N/3$. This follows from the fact that each 
segment is a local minimum with probability $1/3$.
The variance
of $M$ can also be calculated; the result is
\begin{equation}
\mbox{\rm var}(M)= 2N/45 .
\end{equation}
As a consequence the number of local minima $N_{IS}$, which gives
the size of the ISN, is log-normally distributed for
large $N$~\cite{BurdaKrzywicki06} and:
\begin{equation}
\ln \langle N_{IS} \rangle \approx \left[ \frac{1}{3} \ln 2 
+ \frac{1}{45} (\ln 2)^2 \right] N 
\approx 0.2417 N  .
\end{equation}
%

\subsection{Transition links}
\begin{figure}
\includegraphics[width=5cm]{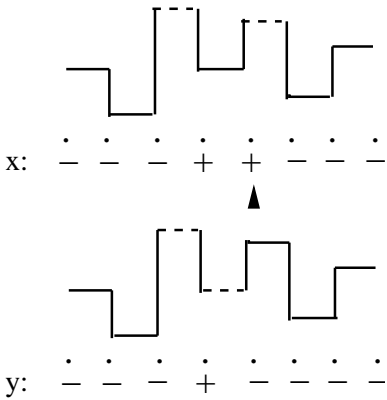}
\caption{\label{fig:transitionState} 
Energy profile representation of a transition state $x$,
and $y$ obtained by flipping the indicated
spin. $Y$ here is the configuration with all
bonds satisfied, while $X$, shown in 
Fig. \ref{fig:landscape}, differs from it
by a droplet consisting of two adjacent 
elementary droplets.}
\end{figure}

Given two inherent structures $X$ and $Y$, 
are they linked by a transition state? This is a priori
a difficult problem as there are many possible
configurations to consider. In 
Fig.~\ref{fig:transitionState} we show how a given
pair $xy$ can be visualized within the 
energy profile picture
in a case where $Y$ is the ground state and $X$ 
has exactly two broken segments, $X$ and $Y$ differing by
two adjacent elementary droplets.
\par
A better understanding of what links two inherent structures
is achieved when one remembers
a few simple facts. First, the set of
unsatisfied bonds determine the system's energy:
if $E_0$ is the ground-state energy where all bonds are
satisfied, then a configuration's energy is given by
\begin{equation}
H(\{\sigma_{i}\}) = E_0 + 2\sum_{i} |J_{i, i+1}|
\end{equation}
where the sum runs over all unsatisfied bonds.
Second, a steepest-descent path corresponds to a specific sequence of
moves where at each step a single spin is flipped
to lower the energy. In such downhill changes,
two alternative possibilities arise. If the flipped spin
belongs to two unsatisfied bonds, then these both become 
satisfied, a phenomenon one can refer to as ``annihilation''.
If on the contrary the flipped spin is shared between
one satisfied bond and one unsatisfied one, then the two
bonds exchange their (satisfied and unsatisfied) nature.
In such a move, the energy must decrease; if we think of
marking broken segments on the energy profile, then the
markings have to go downhill during the 
steepest descent. Note that when two neighboring
unsatisfied bonds can annihilate one another, annihilation
is always preferred to letting one of them go downhill.

In general, the linking of two inherent structures
depends on $\Lambda/J$ and
on the detailed values of the $J_{i,i+1}$; nevertheless
some useful general properties can be derived as follows. 

(i) In the 
energy profile picture, consider a segment $S_{i,i+1}$ which is
a local \emph{maximum}. If the corresponding bond 
is satisfied in both $x$ and $y$
belonging to the basin of $X$ and $Y$ respectively, then all downhill
spin flips will maintain that segment's unbroken state, and thus both $X$ and $Y$
will have that bond satisfied. Thinking of this
in lattice space, let $\sigma_{i_0}$ be the spin that is 
flipped when going from $x$ to $y$.
Starting from $x$ and $y$, the steepest descent energy relaxation
will produce two trajectories of configurations, whose
difference (referred to as ``damage'' in the spin glass literature)
will spread from the initiation site
$\sigma_{i_0}$. 
The corresponding ``damage spreading''~\cite{DerridaWeisbuch87}
front going towards
$i$ will stop at $i$ or before, 
and thus all the spins beyond the ``barrier''
$S_{i,i+1}$ will be blind to this: 
the final difference between $X$ and $Y$
will vanish beyond site $i$ by ``causality''. In effect, information about
the state of $\sigma_{i_0}$ is blocked by this barrier. This same blocking
arises if the maximum segment $S_{i,i+1}$ is broken but 
both trajectories from
$x$ and $y$ flip the same spin on that segment. (Here we use the fact
that of the spins $\sigma_{i}$ and $\sigma_{i+1}$,
one is never flipped while the other is flipped just once 
in the whole steepest descent; indeed, after the first flip, the bond
($i,i+1$) is satisfied and remains so thereafter.)

(ii) Clearly, all maximum segments either start unbroken or
go from broken to unbroken (and remain so) during the steepest
descent moves. The previous causality argument
then shows that the set of maximum segments
that are affected \emph{differently} in the two trajectories from 
$x$ and $y$ cannot have any gaps. Furthermore, since 
in $X$ and $Y$ the spins belonging to the maximum segments
determine completely the spins in the corresponding elementary droplets,
we see that the difference between $X$ and $Y$ must be
due to the flipping of an \emph{uninterrupted} sequence of 
elementary droplets.

\begin{figure}
\includegraphics[width=8cm]{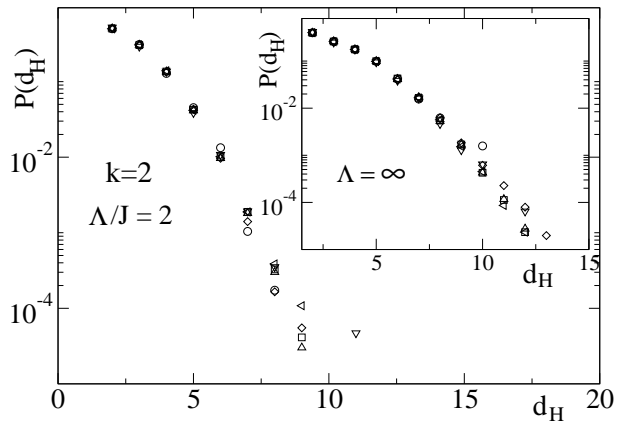}
\caption{\label{fig:OneDimensionCorrelationVolume}
Distribution of the Hamming distance $d_H$ between linked
IS in the one-dimensional lattice for
$\Lambda/J=2$ and for $N$ ranging from $10$ to $30$. 
The average $d_H$ is almost $N$-independent and equals $2.75(2)$.
Inset: same but for $\Lambda=\infty$, the average 
$d_H$ being now $3.20(4)$.}
\end{figure}

Given this last necessary condition for 
inherent structures $X$ and $Y$ to be linked, 
one can also ask whether it is sufficient.
The answer is no:
empirically, it is easy to find disorder samples
(values of the $J_{ij}$) for which no link will
connect $X$ and $Y$ even though they differ by a
connected sequence of elementary droplets.
Indeed, having damage spread
across a series of elementary droplets requires 
conditions on the successive $J_{ij}$.
For example if $J_{1,2}> J_{2,3}> \ldots > J_{m-1,m}$, then the damage
spreading can propagate from site $1$ to site $m$ since
it does not encounter any obstacle in-between, but the probability
that $m$ random numbers form such a decreasing sequence is
rather small: $p(m) \sim 1/m! \sim \exp (-m\ln m)$.
Given these facts, one expects
a probability of propagation that decreases
at least as fast as exponentially with distance. This suggests that
the Hamming distances $d_H$
defined as the number of spins oriented differently in
$X$s and $Y$s have a fast decaying
distribution.
This is indeed what we see in our
numerical simulations, as illustrated in 
Fig.~\ref{fig:OneDimensionCorrelationVolume}.
In fact this fast decay holds for all
$\Lambda$, even $\Lambda = \infty$.
(Note that distances between inherent structures have
also previously been considered in the context
of supercooled liquids in ref.~\cite{FabriciusStariolo02}
and of a one-dimensional Potts model in ref.~\cite{Bertin05}.)
%

\subsection{Degree properties of the ISN}
\begin{figure}[htb]
\includegraphics[width=8cm]{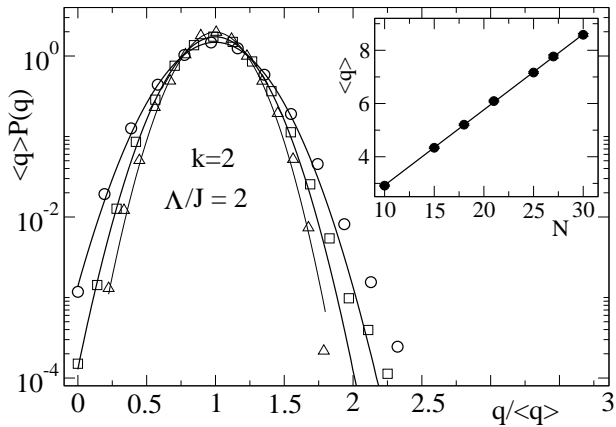}
\caption{\label{fig:oneDimConn} 
Scaled distribution of degree in the ISN
for the one-dimensional lattice
($\Lambda/J=2$) for $N=16$ (circles), $22$ 
(squares), $28$ (triangles) in a semi-logarithmic
plot. The parabola correspond to Gaussian distributions.
Inset: The mean degree versus $N$, along with the 
corresponding linear fit $\langle q\rangle = a + b N$ where
$a = 0.09(3)$ and $b = 0.284(1)$.}
\end{figure}

Let us first discuss some of the topological
properties of the ISN. Consider first the
case $\Lambda/J$ fixed. We saw that the differences (damage) between
linked IS corresponded to a connected cluster of spins
formed by elementary droplets and that the corresponding
distribution of $d_H$ fell sharply. For a given $X$,
the number of such clusters is extensive (proportional to $N$
at large $N$)
and clearly, when they are far away from one another,
they are independent. Thus we expect
IS to have $O(N)$ links. In the inset of Fig.~\ref{fig:oneDimConn}
we show our data for the mean 
degree of inherent structures at $\Lambda/J=2$; the
data agree very well with the 
linear scaling in $N$.
Using the near independence of these clusters, we can also
appeal to the central limit theorem. One then expects
$q$, the degree of an inherent structure, to have a 
Gaussian distribution at large $N$. 
We have corroborated this property also, as illustrated
in the main part of Fig.~\ref{fig:oneDimConn}. Visible deviations
from the Gaussian distribution occur for the small $N$ values,
but go away as $N$ increases.

Now we move on to the case $\Lambda=\infty$.
It is not difficult to see that for any
inherent structure, a link can be formed
by flipping any single one
of the elementary droplets. However it is not
always possible to flip more than that, 
because as was mentioned before, 
there can arise insurmountable barriers to
damage spreading; as a consequence, the number
of elementary droplets that can be jointly flipped
is small. These properties
suggest that the degree properties
of the ISN are not very sensitive to $\Lambda$ 
when $\Lambda$ grows as we now confirm.

\begin{figure}[htb]
\includegraphics[width=8cm]{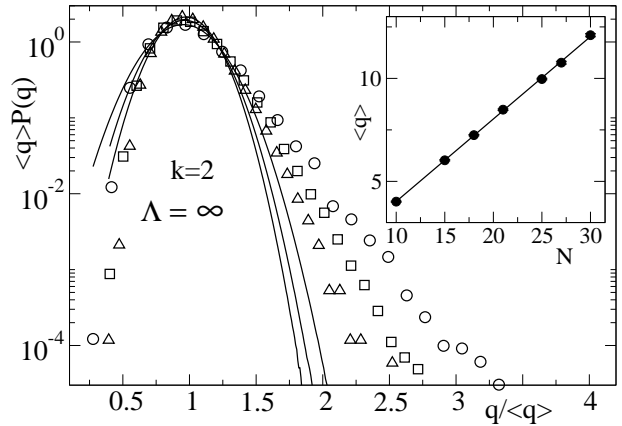}
\caption{\label{fig:oneDimConnbis} 
Scaled distribution of degree in the ISN
for the one-dimensional lattice
($\Lambda=\infty$) for $N=16$ (circles), $22$ 
(squares), $28$ (triangles) in a semi-logarithmic
plot. The parabola correspond to Gaussian distributions.
Inset: The mean degree versus $N$,  along with the 
corresponding linear fit $\langle q\rangle = a + b N$ 
where $a = 0.02(7)$, $b = 0.400(3)$.}
\end{figure}

Consider first the scaling of the mean
degree, $\langle q \rangle$. We see from the 
inset of Fig.~\ref{fig:oneDimConnbis}
that indeed this quantity scales linearly with $N$,
just as in the case of $\Lambda$ finite.
More generally, the slope of the mean 
degree grows with $\Lambda$ but
has a finite limit as $\Lambda \to \infty$.

Second, consider the \emph{distribution} of the degree $q$.
As shown in Fig.~\ref{fig:oneDimConnbis}, 
there is a broad tail at large $q$.
It turns out that the tail comes mostly from links
attached to low energy states:
the collective flipping of multiple elementary droplets is
more likely in that situation.
A subtle correlation, reflecting the shape of the energy profile, 
amplifies that effect: clusters of multiple
elementary droplets can flip more easily when each of their constituents
can flip separately.
\par 
It is very significant that the tail of the distribution shrinks as $N$ 
increases; in particular the effective
slope in the tail gets {\em steeper} with $N$.
Actually, it is not surprising that the deviations from Gaussian
behavior are larger here than when $\Lambda$ is finite: 
the Hamming distances $d_H$ have
a broader distribution 
(cf. Fig.~\ref{fig:OneDimensionCorrelationVolume}).
The tails in Fig.~\ref{fig:oneDimConnbis} arise at frequencies
of $10^{-3}$ or less, and at such frequencies one has
$d_H$ close to 10: to see a clear Gaussian behavior,
it should be necessary to go to $N \gg 10$, and so $N=30$
can be argued to be insufficient.
Given both this theoretical argument and 
the numerical evidence, there is some credence to the claim
that the distribution becomes
Gaussian at large $N$. 

\subsection{Connected components of the ISN}
When $\Lambda/J$ is finite, there may be some bonds whose state
(satisfied or unsatisfied) cannot be changed during the
steepest descent dynamics. Because of this, the ISN will
typically consist of disconnected pieces, where in each
connected component, the spins on such bonds will have fixed
values. 
\par
The existence of such bonds is easily demonstrated:
just consider a bond inside
an elementary droplet; if its height is more than $\Lambda$
above the height of its neighbors, it must always be satisfied
and thus cannot be changed. A consequence of this is that
the spins inside that elementary droplet are frozen over each
separate connected component of the ISN. 
\par
Let us classify
each elementary droplet as being of the ``frozen'' type if 
its orientation cannot be changed when respecting the
bound on energies. If elementary droplets did not touch,
then the frozen nature of an elementary droplet would
not depend on the state of its neighbors; in practice
there is a small dependence, so for instance an elementary
droplet will be frozen in the ground state but not in
some of the excited states. This \emph{dependence on 
context} is weak so it is a good approximation
to consider that elementary droplets
are frozen or not, independently of their surroundings.
An elementary droplet that is not frozen will be called ``active''.

\begin{figure}[htb]
\includegraphics[width=8cm]{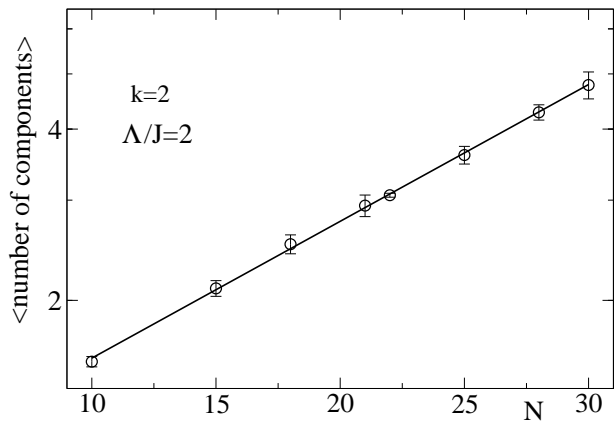}
\caption{\label{fig:numComponents1d} 
Mean number of connected components of the ISN
as a function of $N$ at $\Lambda/J = 2.0$ for the
one-dimensional lattice. Note the
semi-logarithmic scale indicating that the growth
is exponential in $N$. (The best fit 
to the form $\exp{(a + b N)}$ gives
$a = -0.09(1)$ and $b=0.0554(6)$).}
\end{figure}

Within such an ``independent elementary droplets'' approximation,
much can be said about the topology of the ISN.
If there are
a total of $M$ elementary droplets and $(1-f)M$ of these
are frozen, then there will be exactly $2^{(1-f)M}$
components to the ISN. 
(In addition, each of these components
will be isomorphic.)
In Fig.~\ref{fig:numComponents1d}
we confirm this exponential growth in $N$ of the number
of components of the ISN.
\par
Note that the number of frozen elementary droplets
is extensive in $N$, but decreases rapidly as $\Lambda$ grows.
When $\Lambda \to \infty$, all elementary droplets are active
and the ISN is connected. Indeed, to go from any IS to another,
on can simply sequentially turn over
each elementary droplet for which these IS differ: 
a necessary and sufficient condition to have
a connected ISN is for all elementary droplets to be active.

Keeping within this approximation, we can further characterize 
the \emph{structure} of each component of the ISN.
Note that each of the $fM$ active elementary
droplets can be separately flipped, i.e., 
there is one link connecting any IS to the one produced
by flipping any elementary droplet; the degree of each IS will thus
be at least equal to the number of active elementary droplets. 
If these were the only links, each component of the ISN would be
a hypercube of dimension $f M$. But often, 
two inherent structures $X$ and $Y$ will differ by 
\emph{several} consecutive elementary droplets, so each component of
the ISN should be thought of as a hypercube
to which additional links have been added.

A structuring of these additional links arises because 
the (steepest descent) dynamics
occuring inside a region of active elementary droplets delimited
by two frozen elementary droplets is \emph{independent} of what happens outside
of this region. Thus to construct a whole connected component
of the ISN, we can first restrict ourselves to links
coming from multiple elementary droplets lying within a single such
delimited region. 
\par
If there are $p$ elementary droplets
in a delimited region, then we focus on a reduced hypercube of dimension $p$,
the vertices of which are
labeled by the orientation of each of the region's elementary droplets.
The vertices of this hypercube are then connected by links if and only if
a corresponding transition link exists (this can be determined by 
finding transition states, working
\emph{solely} within the delimited region). In addition to the links
between nearest neighbors on the hypercube, there can be longer range links
associated with multiple elementary droplets.
These connections give rise to triangles and other subgraphs
that are absent in the simple picture allowing for independent
two-level systems only. 
\par
Now to get a global description of the ISN,
we use the fact that each IS of the full one-dimensional
lattice is specified by its state in each of the delimited regions.
If there are multiple such regions, then a link between two
IS occurs if and only if they are identical in all but
one region and for that region a link exists as just constructed
on the reduced hypercube.

\section{Generalizing to a framework based on droplets}
\label{sect_DROPLET}

\subsection{Locality and correlation volume}
Elementary droplets played a central role when interpreting the
ISN of the one-dimensional spin glass. In this section we show
that some of the associated concepts form a natural framework
for understanding what can happen in more complex models.
We expose here this more general picture as well as its
predictions, and then shall confront these with 
the actual properties of spin glasses on random graphs.

Consider $\Lambda$ given, $N$ large, and 
a frontier link $xy$; of the steepest
descent paths from configurations $x$ and $y$, at least one
is expected to be composed of $O(\Lambda/J)$ steps
or less because the spin flip at each step generically 
leads to an $O(J)$ decrease of the energy.
(In Fig.~\ref{fig:link},
it is the path from $y$ to $Y$ that is of this type since
it is the one satisfying the $\Lambda$ bound.)
A priori, it is possible that the other path is much longer since 
no bound on the energy applies there, but such a situation
is likely to be rare for typical inherent structures. For 
the sake of simplicity,
we shall neglect this effect and consider that 
both paths have $O(\Lambda/J)$ steps
or less. Since the number of states accessible grows
fast with energy, most of the frontier links should
be close to the bound, and so one can expect to be dominated by
situations in which the number of steps is indeed
$O(\Lambda/J)$, rather than a much smaller number.

Given that each steepest descent path should have $O(\Lambda/J)$ spin flips,
linked inherent structures will differ typically by $O(\Lambda/J)$ spin flips.
Furthermore, the spins of opposite sign in those two IS
cannot be too ``far'' away from one another on the graph $G$. The
reason is causality: when flipping one spin (to go from
$x$ to $y$), the subsequent steepest descent path can undergo
modifications
but only by propagation to nearest neighbor spins at each
step. Thus for a path of $\ell$ steps, only spins
within a distance $\ell$ on $G$ can be affected by
choosing the starting configuration $y$ instead of $x$.
This phenomenon
is in direct correspondence with the standard damage spreading 
dynamics~\cite{DerridaWeisbuch87}.
In the limit of a very large graph $G$ (that is in
the limit $N \to \infty$), for $\Lambda/J$ fixed,
one thus expects the ISN
to have links between $X$s and $Y$s
differing in the orientation of just a few spins; furthermore
these clusters should be \emph{localized} on $G$.
These properties were found to hold very nicely in
the one-dimensional case, though there in addition
the clusters were connected. In general there is no
reason to have only connected clusters, except
in the one-dimensional case where the topology is very special.

To make this picture a bit more quantitative, 
consider the Hamming distance $d_H$
between two linked inherent structures $X$ and $Y$.
Define the mean of $d_H$:
\begin{equation}
v_c \equiv \langle d_H\rangle
\end{equation}
where the average is over linked IS and over samples. One
expects $v_c$ to grow with $\Lambda/J$; the larger $\Lambda/J$
is, the further
the damage spreading from frontier links can propagate.
In effect, $v_c$ describes a kind of
correlation volume. We saw 
in the one-dimensional model that $v_c$ saturates
as $\Lambda \to \infty$ because of the 
presence of insurmountable barriers to damage
spreading in that case; again this phenomenon 
is specific to the topology of the one-dimensional lattice.
On the contrary, we shall see that
$v_c$ diverges when $\Lambda$ becomes unbounded in
the random graph case.

\subsection{Dilute droplets and scaling laws}

Let $\Lambda/J$ be small and take the limit of large $N$. For 
each link of the ISN, we argued that the 
difference between the two corresponding nodes
$X$ and $Y$ should come from a small localized region of spins.
Let us call these ``droplets'' whether or not $X$ and $Y$
can be connected by a barrier less than the \emph{current} $\Lambda/J$.
For simplicity, assume that these droplets are independent;
then if $M$ is the number of droplets in a sample, the number
of inherent structures is $2^M$. Some of these droplets will be ``active''
(have a transition state satisfying the $\Lambda$ bound),
others not. The fraction $f(\Lambda/J)$ of active
droplets grows with $\Lambda/J$. We then see that in this
picture of independent droplets, the ISN consists of
a number of components that grows exponentially in $N$, just 
as we found for the one-dimensional model (cf. Fig.~\ref{fig:numComponents1d}).
In fact, taking the droplet framework literally, the ISN has
$2^{(1-f)M}$ components, each of which is a hypercube of dimension  $f M$;
furthermore, the degree of each node in this ISN is simply $f M$.

$M$ is expected to be extensive (at least
for the kinds of graphs we consider), and to have a Gaussian
distribution when considering samples with different $J_{ij}$.
As a consequence, the number
of inherent structures should have a log-normal distribution,
a property that seems to
hold without restriction~\cite{BurdaKrzywicki06}.
But our droplet framework also predicts that the \emph{degree}
of nodes in the network should have a Gaussian distribution
with a mean scaling linearly with $N$; the coefficient
of this linear scaling is expected to grow
monotonically with $\Lambda/J$.

Of course, these predictions are based on the 
locality argument previously given. As $\Lambda/J$
grows, more droplets arise, their $v_c$ also grows,
and so the independent (dilute)
approximation may break-down. Although
no such break-down occurs for the one-dimensional
model, we shall see that in the case of 
$k=4$ regular random graphs, the system's behavior
changes dramatically when $\Lambda \to \infty$.

\section{The random graph case}
\label{sect_INFINITEDIMENSION}
\subsection{Qualitative aspects}
We now move on to the case where the spins lie on sites
belonging to a connected $k$-regular random graph, one of
the standard frameworks for mean-field studies of
spin glasses. The connectivity property is desirable
because we are limited to rather few spins; it
would be unreasonable to allow our small
systems to be made up of even smaller
independent subsystems. In this class
of graphs, each of the $N$ sites is connected to exactly
$k$ other sites, with an associated $J_{ij}$,
the Hamiltonian being given by Eq.~(\ref{eq:H}).
We have chosen $k=4$ to have a value neither too small
nor too large. Indeed, $k=3$ random graphs
have a non negligible probability of being disconnected
for the $N$ values we use
(the connectivity can be enforced, but it is simpler to
set $k=4$; with this choice randomly generated 
graphs are in practice never disconnected).
On the other hand,
as $k$ grows, the consequences of sparseness set in at
larger and larger $N$ values, and since we are limited
to $N \le 30$, it is preferable to avoid this.

These graphs have two types of quenched disorder, coming from
the couplings ($J_{ij}$) and from the ``geometry''
(which sites are coupled to which). In comparison to the
one-dimensional case, there are several very important
differences associated with the graph's more complex topology.
First, one cannot gauge away the signs of the $J_{ij}$
to make all couplings positive: the system is inherently
frustrated. Second, one cannot introduce a local
structure that will automatically block damage spreading;
any obstacle can be ``by-passed'', a feature that (at least
for nearest-neighbor interactions) has
no analog in one dimension. Third, elementary droplets
cannot be defined a priori, and no simple characterisation
of inherent structures allows one to enumerate
them efficiently. The reason is that
droplets are context sensitive. Specifically, this means that
if one takes the cluster of spins
defining a droplet between inherent structures $X$ and $Y$, 
flipping that same cluster in another inherent structure
$Z$ will not in general lead to a local minimum of the
Hamiltonian. 

Because droplets are context sensitive, the number of inherent structures
$N_{IS}$ will generally not be a power of $2$. 
Nevertheless, one expects there will be many local excitations
and this should be enough to make 
$N_{IS}$ have a log-normal distribution. This 
indeed seems to be the case as has been found before, 
in particular in~\cite{BurdaKrzywicki06}.

\subsection{Droplet sizes}
Let us now focus on the Hamming distance
$d_H$ between linked inherent structures. To begin,
assume that $\Lambda$ is fixed.
\begin{figure}[htb]
\includegraphics[width=8cm]{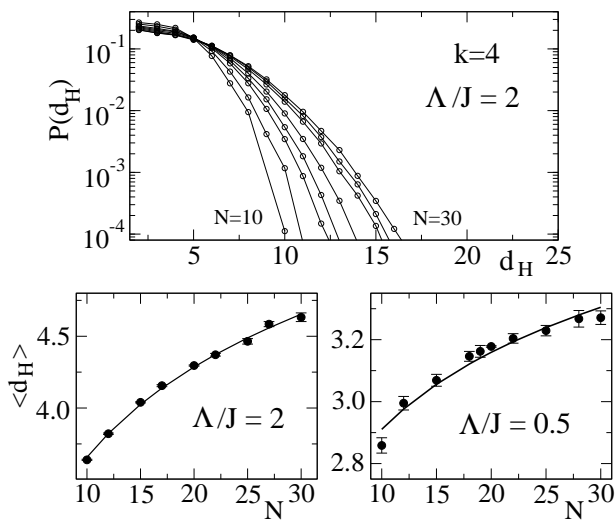}
\caption{\label{fig:mfDistvc} 
Top: the distribution
of the Hamming distance between linked
IS  for $N$ from $10$ to $30$ at $\Lambda/J=2$. 
The lines are to guide the eye. Below: The 
average Hamming distance versus $N$
for $\Lambda/J=2$ (left) and $\Lambda/J=0.5$ 
(right), along with logarithmic fits.}
\end{figure}
In the one-dimensional case, droplets were localized
and connected, with $d_H$ having a distribution falling
off fast at large values and almost no $N$ dependence.
We show in Fig. \ref{fig:mfDistvc} what happens in 
the random graph case. At the top of that figure
we show the distribution when 
$\Lambda/J=2$, for $N$ values ranging from $10$ to $30$.
One sees that the distribution of $d_H$ still falls off relatively
fast, perhaps faster than an exponential, just as in 
the one-dimensional case. However there is now
a trend in $N$, the droplets typically growing
when $N$ increases. In the bottom of 
Fig.~\ref{fig:mfDistvc} we show $v_c \equiv \langle d_H \rangle$
as a function of $N$.
For $\Lambda/J=2$ the data may saturate at large $N$, but are 
also compatible with a logarithmic growth.
(The curve on the left is $v_c = 1.57(5) + 0.91(2) \ln(N)$). 
Since this growth could be just a finite size effect,
we also show a similar plot but at a smaller value
of $\Lambda/J$, viz. $\Lambda/J=0.5$. Here the saturation
seems more likely: the logarithmic fit is quite
poor while a fit to the form $a + b/N$ is 
very good. Note that random graphs naturally
produce $1/N$ effects because the probability that
a given site belongs to a small loop (say a triangle) is $O(1/N)$.
We conclude that the observed growth is plausibly
a finite-size effect although a growth continuing
indefinitely at higher $N$ cannot be completely excluded.

One might guess that the divergence of $v_c$ with $N$ depends
on the value of $\Lambda$; however, our claim is that on the 
contrary all finite $\Lambda$ values lead
to the same behavior: either $v_c$ always diverges when $N\to \infty$
or it never does.
The reasoning is as follows. For specificity, take the
situation illustrated in Fig.~\ref{fig:link} where
$x$ satisfies the bound with respect to $Y$. The configurations
$x$ and $y$ differ from $Y$ by a finite number of spin flips.
Let us assume that $v_c$ diverges so there
is a non-zero probability 
for the damage spreading front to go arbitrarily far.
In that ``far away'' regime, the front
advances in a region of the graph $G$ where
the spins are far
from the few spins that are flipped to go from
$Y$ to $x$. This ``invasion
process'' will initiate with a probability that grows with
$\Lambda$, but because the $J_{ij}$ are continuous
and the associated local field on each spin has a finite
density at $0$, this initiation happens with a strictly
positive probability at all $\Lambda$.
The bound $\Lambda$ just plays the role of limiting
the initiation probability, but does not affect
the ability of the invasion process to spread arbitrarily far.
Thus if $v_c$ diverges for one value of $\Lambda$,
then it must diverge for all other finite
values of $\Lambda$. (It may be possible to
use the techniques developped in ref.~\cite{MontanariSemerjian06}
to compute properties of this invasion process,
at least on certain kinds of graphs.)
To summarize, it seems we have just the following 
two alternative possibilities:
(i) $v_c$ diverges with $N$, possibly logarithmically;
(ii) $v_c$ saturates as $N \to \infty$.
In both cases the behavior holds for all finite values of $\Lambda$.

Coming back to our data, we see that in practice
damage spreading remains \emph{relatively} localized. Note
that droplets need not be connected and sometimes
indeed are not, but that is rather exceptional.
Comparing these results to the 
case of the one-dimensional spin glass, we see that
the droplets are definitely larger here. Probably this has
two sources: first no single
obstacle can stop damage spreading, and
second the high level of frustration in the
present model should go hand in hand with 
enhanced fragility to perturbations.
Because of these larger droplets, one has to go to larger
$N$ to hope to see the large $N$ scaling set in. 
In particular, for sure one needs
the total number of spins $N$ to be much larger than the
mean droplet volume $v_c$.
However, it might be argued that it is the diameter of the 
graph that should be much larger than the diameter of the droplets;
since the graph's diameter grows as $\ln (N)$, 
this would imply going to much larger values
of $N$ to see the scaling set in convincingly. Such 
large sizes are way beyond the reach of current techniques.
Note that the near logarithmic growth of $v_c$ with 
$N$ seen at $\Lambda=2.0$ may be related to the fact that
the the graph's diameter grows as $\ln (N)$, the scale
where loops set in.

Finally, consider the case $\Lambda=\infty$. In the one-dimensional
model, insurmountable barriers prevented $v_c$ from growing much,
but here there is no reason for $v_c$ not to diverge as
$\Lambda \to \infty$. Our data for $k=4$ random graphs
are unambiguous here: we
observe a very clean linear growth of 
$v_c$ with $N$, i.e., damage spreading can invade the whole system.
Thus in random graph case (in contrast to what happens in one dimension),
the droplet framework will be of no use for interpreting
properties of the ISN when $\Lambda=\infty$: not only
are droplets strongly correlated there, they are also delocalized.

\subsection{Degree properties at $\Lambda=\infty$}
Let us first investigate the case $\Lambda=\infty$. As mentioned before, 
when energies are unbounded, locality no longer holds and so
the droplet framework is misleading.

\begin{figure}[htb]
\includegraphics[width=8cm]{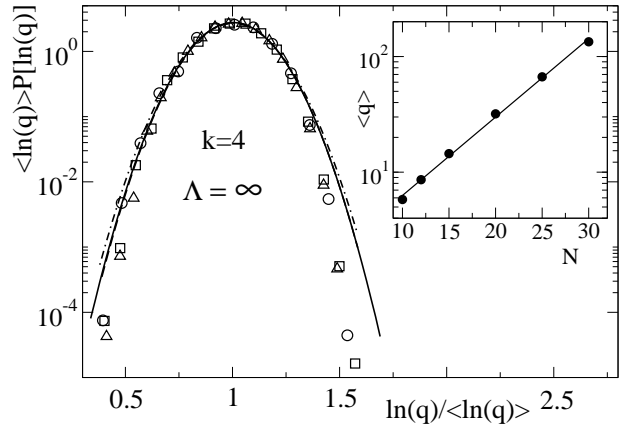}
\caption{\label{fig:conndistrinf} 
Scaled degree distribution at $N=20$ (circles), $25$ (squares)
and $30$ (triangles) for $\Lambda=\infty$.
$P(\ln q)$ is normalized to unity; the bin size $\Delta\ln{(q)}= 0.3$. 
The curves are obtained from log-normal fits to the data.
In the inset: average degree versus $N$ (notice the logarithmic 
vertical scale; the best fit is 
$\langle q\rangle = \exp{(a + b N)}$ where
$a=0.28(8)$ and $b=0.156(4)$).}
\end{figure}

Consider a configuration $x$ of high energy
which under steepest descent goes to the 
inherent structure $X$. At high energies,
damage spreading is expected to be able propagate far away, so that
$X$ should be linked to many $Y$ whose Hamming distance is
proportional to $N$. Since there are no insurmountable barriers to 
damage spreading, the spreading can be rather sensitive
to details of the configuration $x$; this picture suggests
that each IS is linked to many others on the ISN, presumably
an exponentially large number with $N$. This is borne
out by our simulations. If $q$ is the number of links
attached to $X$, we find that $\langle q \rangle$
grows exponentially with $N$ as illustrated in 
the inset of Fig.~\ref{fig:conndistrinf}; our best fit gives
\begin{equation}
\langle q\rangle = e^{a + b N} \,\, \mbox{\rm with} \,\,
a=0.28(8) \, , \, b=0.156(4) .
\end{equation}

We know that $N_{IS}$, the number of inherent
structures, is exponentially large with $N$,
and here we find that $\langle q \rangle$ is also, 
but nevertheless the mean degree of the ISN is 
much smaller than the maximum possible value. Comparing 
with the data of ref.~\cite{BurdaKrzywicki06}
one finds that 
\begin{equation}
\langle q\rangle \propto \langle N_{IS} \rangle^\beta
\end{equation}
where
the exponent $\beta \approx 0.72$ is significantly less than unity. 
\par
We also
mentioned that the number of IS has a log-normal
distribution; the argument behind that can plausibly
be applied to $q$ itself. Thus in Fig.~\ref{fig:conndistrinf}
we compare the empirical distribution of $q$ to 
a log-normal one adjusted to have the same
mean and variance. The data indicate
that the distribution is steeper than log-normal,
but also less steep than a Gaussian. We also find that 
this distribution is relatively insensitive to the 
$N$ values studied. But this may be 
misleading: if we consider
the variance of $\ln{q}$, it increases roughly
like $\langle \ln{q}\rangle$, though the finite-size
effects are sizeable. If one 
extrapolates the behavior of the largest system sizes accessible to us,
namely $20 \le N \le 30$,
one finds that the variance of the quantity
$\ln{q}/\langle \ln{q}\rangle$, namely
$\mbox{\rm var}(\ln{q})/\langle \ln{q}\rangle^2$ can be fit to the form
\begin{equation}
\frac{\mbox{\rm var}(\ln{q})}{\langle \ln{q}\rangle^2} \sim
\frac{a}{N}(1+b/N+...)
\end{equation}
with $b \approx -10$. This coefficient is large, making the extrapolation
dangerous, but if it is correct, 
the distribution at large $N$ is log-normal, so
the bulk of the scaled degree distribution shown 
in Fig.~\ref{fig:conndistrinf}
will shrink, but only for $N \gg 10$. 

\subsection{Degree properties at $\Lambda$ finite}
Now we come to the case of $\Lambda$ finite, where
we found droplets to be typically localized, and so the droplet
framework may be expected to be a useful guide.

\begin{figure}[htb]
\includegraphics[width=8cm]{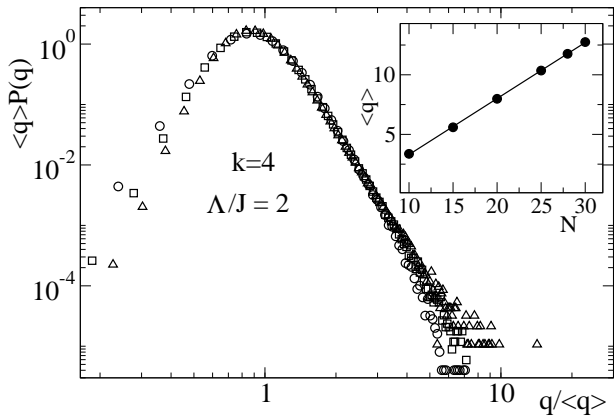}
\caption{\label{fig:conndistr2} 
Scaled degree distribution at $N=20$ (circles), $25$ (squares)
and $30$ (triangles) for $\Lambda/J = 2$.
$P(q)$ is normalized to unity. 
In the inset: average degree versus $N$ (the best 
fit of the form $\langle q\rangle  = a + b N$ gives
$a= -1.40(6)$ and $b=0.470(3)$.} 
\end{figure}

The first prediction of the droplet framework is that
$\langle q \rangle$ grows linearly with $N$.
Our data are in good agreement with this scaling, and for
illustrative purposes we display in the inset of
Fig.~\ref{fig:conndistr2} the linear growth of $\langle q \rangle$ with
$N$ when $\Lambda/J=2$; our best linear fit leads to 
\begin{equation}
\langle q\rangle  = a + b N \,\, \mbox{\rm with} \,\,
a= -1.40(6) \, , \, b=0.470(3). 
\end{equation}

\begin{figure}[htb]
\includegraphics[width=8cm]{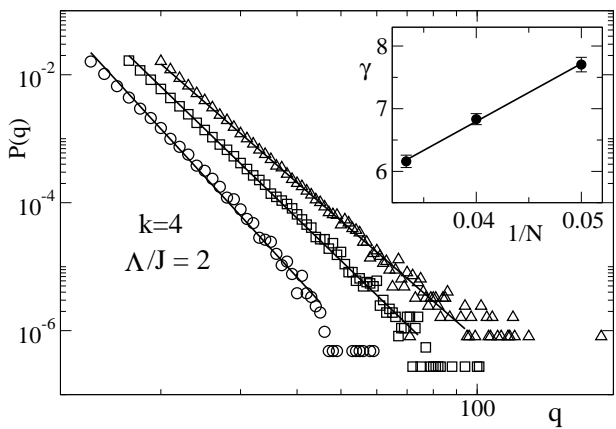}
\caption{\label{fig:fittail} 
Tails of the degree distributions at $N=20$ (circles), $25$ (squares)
and $30$ (triangles) for $\Lambda/J = 2$.
$P(q)$ is normalized to unity. 
Power fits $P(q) \propto q^{-\gamma}$ are also shown.
In the inset: the exponent $\gamma$ versus $1/N$; a very tentative linear 
fit yields $\gamma = 3.1(2) + 92.2(3.8)/N$ (notice that the 
origin of the horizontal axis is far to the left!).}
\end{figure}

The second prediction of the droplet framework is
that $q$ has a Gaussian limiting distribution.
The main part of Fig.~\ref{fig:conndistr2} suggests
on the contrary that the distribution has a power law
tail at large $q$.
There is no indication that the fat tail
observed is a finite $N$ effect that goes
away as $N$ increases: a zoom on the tails
for different $N$ shows this, as 
is displayed in Fig.~\ref{fig:fittail}.
At each $N$ the tail is compatible
with a pure power law; the associated exponent 
tends to less negative values with increasing $N$.
We have performed a 
(tentative) $1/N$ extrapolation on the power and this 
works quite well (see inset), leading to the estimation
of an algebraic decay with exponent $\approx 3$ 
in the large $N$ limit. All of these properties
hold for the different values of $\Lambda$
we have investigated; in particular the exponents
of the tails are insensitive to $\Lambda$.
\par
The behavior of the distribution is striking, because 
in the droplet framework the quantity
$q/\langle q \rangle$ is expected to become peaked
whereas we see instead a stable distribution
with a fat tail. If this is indeed indicative
of the large $N$ behavior, then $q$ is not self-averaging; such 
a situation is surprising and requires strong correlations,
and presumably a diverging $v_c$. Distributions with 
power law tails are often called ``scale-free'', and
many natural and artificial systems have networks with 
this property. Growth rules have been proposed to
explain the relative ubiquity of scale-free 
networks~\cite{BarabasiAlbert99,GuimaraesdeAguiar05},
but our ISN is not the result of a growth process; thus
other explanations are called for.

Just as we did for $\Lambda=\infty$, 
one may ask whether there is any trend that would
suggest a narrowing of the distribution with increasing $N$.
We have thus examined the variance
of $q/\langle q \rangle$, namely
$\mbox{\rm var}(q)/\langle q\rangle^2$. It is possible to fit it
to the form
\begin{equation}
\frac{\mbox{\rm var}(q)}{\langle q\rangle^2} \sim
\frac{a'}{N}(1+b'/N+...)
\end{equation}
with $b' \approx -10$. Again this correction term is large
and compensates to a large extent the trend imposed
by the first term; it also shows that a putative
narrowing of the distribution 
can only set in for $N \gg 10$. However, in such
a picture, one would still expect
the amplitude of the tails in Fig.~\ref{fig:conndistr2} 
to diminish with $N$, while a close observation
of that figure shows that they grow slightly.

To conclude, we can interpret the data either as
giving some credence to the droplet claim
whereby a central limit behavior will transpire
but only for much larger $N$ than we can tackle,
or, more likely, as
giving evidence for a limiting distribution for
$q/\langle q \rangle$ at large $N$ with power law tails.
In this last scenario, the system is ``critical''
for all finite $\Lambda$, suggesting that droplets
are correlated on all scales. Since no parameter 
has been fine tuned, the criticality
is self-organized~\cite{BakChao88}.

\subsection{Connected components of the ISN}
\begin{figure}[htb]
\includegraphics[width=8cm]{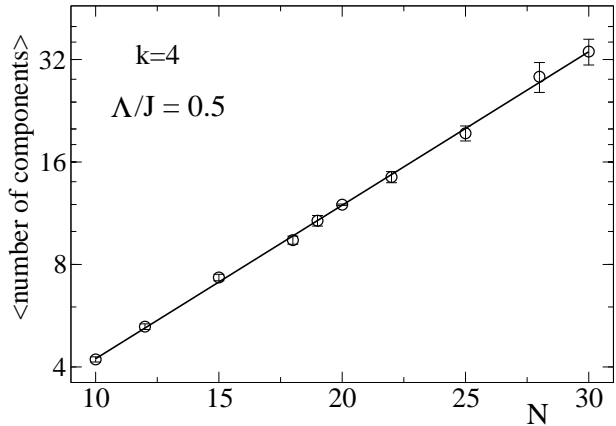}
\caption{\label{fig:numComponentsMF} 
Mean number of connected components of the ISN
as a function of $N$ at $\Lambda/J = 0.5$ for the
spin glass on 
($k=4$ regular) random graphs.
Note the semi-logarithmic scale indicating that the growth
is exponential in $N$ (the best fit is $\exp{(0.40(3)+ 0.104(2) N)}$).}
\end{figure}

Just as in the one-dimensional case, it is of interest to
understand the connectivity properties of the ISN. Since
in the random graph case, damage spreading is less subject to bottlenecks,
one expects the ISN to have fewer components than
in the one-dimensional case. The greater
fragility of configurations to perturbations also suggests this.
Not surprisingly, 
our simulations confirm these expectations. For instance, when $\Lambda/J=2$, 
the one-dimensional case typically had multiple components, 
whereas in the random graph case usually there
is just one component. To obtain more components, it is necessary
to go to smaller values of $\Lambda/J$. In Fig.~\ref{fig:numComponentsMF}
we show how the mean number of components 
grows with $N$ for $\Lambda/J=0.5$~:
we find again an exponential increase,
the best fit being $\exp{(0.40(3)+ 0.104(2) N)}$.

This qualitative behavior can
be understood as follows. Even though there are no
``topological'' barriers in the spin glass model defined
on random graphs, the set of bonds whose couplings $J_{ij}$ have 
particularly large magnitudes can disconnect the ISN. Indeed,
when $\Lambda/J$ is too small to allow for unsatisfying 
a bond, then that bond is frozen in an entire connected
component of the ISN. The number of these frozen bonds
grows extensively with $N$, and so the number of components
is expected to grow exponentially with $N$.
As $\Lambda/J$ grows, the number of frozen bonds drops
very fast, so for relatively modest values
of $\Lambda$ (given our small values of $N$)
the ISN has just a single connected component.

\section{Discussion and conclusion}
\label{sect_CONCLUSIONS}
\par
For studying the cooperative behavior of complex systems,
it is common practice to refer to inherent 
structures~\cite{StillingerWeber84,AngelaniParisi98,
SciortinoKob99,Shastry01}. Here we took model systems 
based on spin-glass Hamiltonians and determined the 
degree and connectivity properties of the ISN, the network which 
interconnects the inherent structures. That network 
architecture is important for the system's dynamical behavior. 
\par
This kind of study has been pioneered in Refs.~\cite{Doye02,DoyeMassen05}
in a context where the configuration space is continuous. In the
models we studied here, this space (a Boolean hyper-cube) is discrete. Thus
our first task has been to introduce an appropriate definition of
the transition states, the passes between two neighboring basins 
of attraction of two distinct inherent structures
(see Fig.~\ref{fig:link}). We found useful 
to include in this definition a cut-off $\Lambda$: 
the energy of a transition state cannot be more than 
$\Lambda$ above both of its inherent structures.
Of course, when $\Lambda/J$ is very small, the ISN is very weakly 
connected and of little interest. We have found, however, that 
for $\Lambda/J$ of order unity one observes in addition to
the well known asymptotics for the number
of inherent structures
\begin{equation}
\langle N_{IS}\rangle \propto e^{\alpha N}
\end{equation}
the following characteristic scaling behavior
of the average number of transition 
states $N_T$~: 
\begin{equation}
\label{scaling}
\langle N_T \rangle \propto N e^{\alpha N}
\end{equation}
as would hold in a system regarded as a 
collection of equivalent and independent 
subsystems~\cite{StillingerWeber82,DoyeWales02,Wales03}. 
(Notice that this implies that the average degree of the ISN is 
linear in $N$). The scaling law (\ref{scaling}) is likely to hold in 
the thermodynamic limit, for any finite value of $\Lambda/J$, but we 
have no proof that it is indeed so. In any case, Eq.~(\ref{scaling}) 
suggests to interpret the data in terms of local excitations. This 
is the essence of our approach, our ``droplets'' are the 
local excitations of the system.
\par
The droplet framework in its simplest version emerges naturally when
the spins live on a one-dimensional lattice. In this case, for
every given $\{J_{ij}\}$, the enumeration of inherent structures is
particularly simple: the lattice can be unambiguously divided into
non-overlapping intervals and in each interval the spins are either
all $+1$ or all $-1$ (in an appropriate gauge). Thus the inherent 
structure network is essentially based on independent two-level systems,
the elementary ``droplets'' formed of 
connected clusters of spins.
\par
Although the inherent structures can easily be visualized and counted
for the one-dimensional spin glass,
the inherent network is nevertheless non trivial since it is not obvious
at all which metastable states are connected via transition
states. In the simplest case just the spins of an elementary droplet 
are flipped. A flip of \emph{several} elementary droplets is
also possible, under certain conditions: a necessary one is 
that these droplets form an uninterrupted sequence.
These elementary droplets are de facto
correlated, reflecting
the shape of the ``energy profile'' of bonds: it turns out that the
likelihood of a flip of a ``compound'' droplet made up of $n$
elementary ones is enhanced when flips of spins in ``smaller''
droplets made of $m<n$ elementary ones are likely.
This effect produces a tail in the degree distribution.
However, the topology of our one-dimensional lattice implies that 
the correlations effectively involve on average
only a finite number of droplets.
Consequently, Eq.~(\ref{scaling}) holds for all $\Lambda$. Furthermore,
the degree of an ISN node is a sum of a number of independent
random variables and the degree distribution tends to a Gaussian
as $N \to \infty$.
\par
The droplet framework being an excellent guide for the
one-dimensional spin glass, it is natural to extend it to
the more complicated mean-field spin glass
defined on random graphs, although in this case the
droplets are not defined without reference to the 
underlying inherent structures. If the one-dimensional lattice is
represented by a ring, then a 4-regular graph can be regarded as
a ring with multiple short-cuts; this changes the way
droplets can interact and their locality is no longer evident.
Furthermore, the properties of the model depend qualitatively on 
whether $\Lambda$ is finite or not, and in practice
on the magnitude of $\Lambda$.
\par
When $\Lambda/J \gg 1$, the scaling property, Eq.~(\ref{scaling}), no longer 
holds and the correlation volume (or mean droplet
size) becomes comparable to the whole system. 
As a consequence, the droplet framework is of no use. For this strongly
correlated system, instead of Eq.~(\ref{scaling}),
we find that
the mean degree of the ISN grows exponentially with $N$. Finally one
can argue that the degree distribution should
be log-normal and this is relatively compatible with our data.
\par
On the contrary, when $\Lambda/J$ is of order unity, the picture is
closer to that of the droplet framework
except that the finite-size
corrections are large. In particular,
most properties we find can again be interpreted in the droplet framework, with
one notable exception: for $\Lambda/J$ of order unity, the 
degree distribution has a scale-free tail,  
$P(q) \propto q^{-\gamma}$ for large $q$. Moreover, and this
is particularly significant, the exponent $\gamma$ {\em decreases}
with increasing $N$. A tentative (and bold!) extrapolation suggests
that it might be close to 3 in the thermodynamic limit. We did not
succeed to find a plausible explanation of this phenomenon. Clearly
random graphs as used in mean-field spin glasses
have strongly correlated droplets, allowing for collective
flipping of spin clusters in the range for $N$ accessible
to our study. If these correlations are so strong as to
maintain a fat tail distribution in the thermodynamic
limit, then clearly the droplet framework is inadequate
and instead some self-organizing
criticality~\cite{BakChao88} principle is at work.
\par
Our approach differs from that used in the literature
for certain atomic clusters~\cite{Doye02,DoyeMassen05}, where
the ISN structure has been interpreted using
geometric arguments. We did not follow that avenue for
several reasons.  First, the idea that the scale-free tail of
the degree distribution might reflect the properties of
a dense packing of basins does not work for the system
studied here: if it were true one would a fortiori observe this
feature at $\Lambda=\infty$, when all neighbor basins are
connected, but this does not happen. Second, the idea that
the partition of the configuration space into attraction basins
exhibits a fractal structure is untenable in
our context: in our configuration space, i.e., on the Boolean
cube, the number of points located at Hamming distance $r$ from 
a given vertex is ${N \choose r}$, which for $r \ll N$ equals
$N^r/r!$. This shows that the space has a negative curvature, most
points of a basin are located near its boundary. This is like in a
symmetric Cayley tree and explains why a basin
can have so huge a number of neighbors: the boundary itself is huge.
Now, self-similarity in a curved space is a somewhat ill
defined concept, because the curvature sets a distance
scale. Finally, we measured the basin size distribution; in contrast to 
the systems studied in refs.~\cite{Doye02,DoyeMassen05}, it
{\em is not} scale-free.
\par
In spite of a different way of looking at 
our systems, we recover
several of the geometric observations made in refs.~\cite{Doye02,DoyeMassen05}.
The hubs of the ISN are due to low energy states,
in fact often the ground state for the $N$ values
we can tackle in this work.
And there is a positive correlation between
the degree of an ISN node and the size of the corresponding
basin. 

When considering the connectivity properties of ISN,
we showed that for finite $\Lambda$ the number of 
components grew exponentially with $N$.
We argued that this was necessarily the case when
values of the $J_{ij}$ could be larger
than $\Lambda$ since the associated bonds have
to remain satisfied throughout the whole steepest
descent from the transition states. Each such bond
breaks the different components of the ISN into two
further pieces that are
of nearly identical size. As a consequence,
at any fixed $\Lambda$, the largest component of the ISN
represents an exponentially small fraction of
the whole as $N\to \infty$: because of this there is no 
percolation transition.
Our model can thus be contrasted with the
continuous energy landscape model of Weinrib and 
Halperin~\cite{WeinribHalperin82}:
there, by increasing the barrier value
in the thermodynamic limit, a 
true percolation transition is found.

\par
As already stated, we have no explanation of the
appearance of the scale-free tail in the degree distribution:
it remains a bit mysterious. The standard de-excitation of a
metastable state occurs via a cascade of small steps connecting
energetically close levels (e.g. in quantum mechanics this is 
due to the direct relation between transition probabilities and
wave function overlaps). In a complex system like the one 
studied here there is a significant probability that many
excited states are separated by a small
energy barrier from one low energy state, which then appears to be
a hub of incoming links on the ISN. Manifestly, this is not an 
uncommon feature, since it has been observed in a set
of very different systems. One can wonder what use 
Nature can make of this curious geometry. A clarification of this
issue remains an interesting challenge.

\par
{\bf Acknowledgments:}
We are very indebted to B. Waclaw for his 
contribution to the improvement of the 
efficiency of our numerical code. We also thank
Satya Majumdar for helpful comments.
This work was supported by the EEC's FP6 Information
Society Technologies Programme
under contract IST-001935, EVERGROW (www.evergrow.org), 
by the EEC's FP6 Marie Curie RTN under
contract MRTN-CT-2004-005616 (ENRAGE: European
Network on Random Geometry), by the
Marie Curie Actions Transfer of Knowledge project ``COCOS'',
Grant No. MTKD-CT-2004-517186 and by the Polish Ministry of Science
and Information Society Technologies Grant 1P03B-04029 (2005-2008).
The LPT and LPTMS are Unit\'e de Recherche de
l'Universit\'e Paris-Sud associ\'ees au CNRS.

\bibliographystyle{apsrev}

\bibliography{references,landscapes}

\addcontentsline{toc}{chapter}{\protect\bibname}\addcontentsline{toc}{chapter}%
{\protect\bibname}
\begin{thebibliography}{32}
\expandafter\ifx\csname natexlab\endcsname\relax\def\natexlab#1{#1}\fi
\expandafter\ifx\csname bibnamefont\endcsname\relax
  \def\bibnamefont#1{#1}\fi
\expandafter\ifx\csname bibfnamefont\endcsname\relax
  \def\bibfnamefont#1{#1}\fi
\expandafter\ifx\csname citenamefont\endcsname\relax
  \def\citenamefont#1{#1}\fi
\expandafter\ifx\csname url\endcsname\relax
  \def\url#1{\texttt{#1}}\fi
\expandafter\ifx\csname urlprefix\endcsname\relax\def\urlprefix{URL }\fi
\providecommand{\bibinfo}[2]{#2}
\providecommand{\eprint}[2][]{\url{#2}}

\bibitem[{\citenamefont{{Stillinger} and {Weber}}(1984)}]{StillingerWeber84}
\bibinfo{author}{\bibfnamefont{F.~H.} \bibnamefont{{Stillinger}}}
  \bibnamefont{and} \bibinfo{author}{\bibfnamefont{T.~A.}
  \bibnamefont{{Weber}}}, \bibinfo{journal}{Science}
  \textbf{\bibinfo{volume}{225}}, \bibinfo{pages}{983} (\bibinfo{year}{1984}).

\bibitem[{\citenamefont{B\"uchner and Heuer}(1999)}]{BuchnerHeuer99}
\bibinfo{author}{\bibfnamefont{S.}~\bibnamefont{B\"uchner}} \bibnamefont{and}
  \bibinfo{author}{\bibfnamefont{A.}~\bibnamefont{Heuer}},
  \bibinfo{journal}{Phys. Rev. E} \textbf{\bibinfo{volume}{60}},
  \bibinfo{pages}{6507} (\bibinfo{year}{1999}).

\bibitem[{\citenamefont{Bogdan et~al.}(2006)\citenamefont{Bogdan, Wales, and
  Calvo}}]{BogdanWales06}
\bibinfo{author}{\bibfnamefont{T.}~\bibnamefont{Bogdan}},
  \bibinfo{author}{\bibfnamefont{D.}~\bibnamefont{Wales}}, \bibnamefont{and}
  \bibinfo{author}{\bibfnamefont{F.}~\bibnamefont{Calvo}}, \bibinfo{journal}{J.
  Chem. Phys.} \textbf{\bibinfo{volume}{124}}, \bibinfo{pages}{044102}
  (\bibinfo{year}{2006}).

\bibitem[{\citenamefont{Crisanti et~al.}(2000)\citenamefont{Crisanti, Ritort,
  Rocco, and Sellitto}}]{CrisantiRitort00}
\bibinfo{author}{\bibfnamefont{A.}~\bibnamefont{Crisanti}},
  \bibinfo{author}{\bibfnamefont{F.}~\bibnamefont{Ritort}},
  \bibinfo{author}{\bibfnamefont{A.}~\bibnamefont{Rocco}}, \bibnamefont{and}
  \bibinfo{author}{\bibfnamefont{M.}~\bibnamefont{Sellitto}},
  \bibinfo{journal}{J. Chem. Phys.} \textbf{\bibinfo{volume}{113}},
  \bibinfo{pages}{10615} (\bibinfo{year}{2000}).

\bibitem[{\citenamefont{Doye and Wales}(2002)}]{DoyeWales02}
\bibinfo{author}{\bibfnamefont{J.~P.~K.} \bibnamefont{Doye}} \bibnamefont{and}
  \bibinfo{author}{\bibfnamefont{D.~J.} \bibnamefont{Wales}},
  \bibinfo{journal}{J. Chem. Phys.} \textbf{\bibinfo{volume}{116}},
  \bibinfo{pages}{3777} (\bibinfo{year}{2002}).

\bibitem[{\citenamefont{Giovambattista
  et~al.}(2002)\citenamefont{Giovambattista, Starr, Sciortino, Buldyrev, and
  Stanley}}]{GiovambattistaStarr02}
\bibinfo{author}{\bibfnamefont{N.}~\bibnamefont{Giovambattista}},
  \bibinfo{author}{\bibfnamefont{F.}~\bibnamefont{Starr}},
  \bibinfo{author}{\bibfnamefont{F.}~\bibnamefont{Sciortino}},
  \bibinfo{author}{\bibfnamefont{S.}~\bibnamefont{Buldyrev}}, \bibnamefont{and}
  \bibinfo{author}{\bibfnamefont{H.}~\bibnamefont{Stanley}},
  \bibinfo{journal}{Phys. Rev. E} \textbf{\bibinfo{volume}{65}},
  \bibinfo{pages}{041502} (\bibinfo{year}{2002}).

\bibitem[{\citenamefont{Wales and Doye}(2003)}]{DoyeWales03}
\bibinfo{author}{\bibfnamefont{D.~J.} \bibnamefont{Wales}} \bibnamefont{and}
  \bibinfo{author}{\bibfnamefont{J.~P.~K.} \bibnamefont{Doye}},
  \bibinfo{journal}{J. Chem. Phys.} \textbf{\bibinfo{volume}{119}},
  \bibinfo{pages}{12409} (\bibinfo{year}{2003}).

\bibitem[{\citenamefont{Doliwa and Heuer}(2003)}]{DoliwaHeuer03}
\bibinfo{author}{\bibfnamefont{B.}~\bibnamefont{Doliwa}} \bibnamefont{and}
  \bibinfo{author}{\bibfnamefont{A.}~\bibnamefont{Heuer}},
  \bibinfo{journal}{Phys. Rev. E} \textbf{\bibinfo{volume}{67}},
  \bibinfo{pages}{030501} (\bibinfo{year}{2003}).

\bibitem[{\citenamefont{Stariolo et~al.}(2004)\citenamefont{Stariolo, Arenzon,
  and Fabricius}}]{StarioloArenzon04}
\bibinfo{author}{\bibfnamefont{D.}~\bibnamefont{Stariolo}},
  \bibinfo{author}{\bibfnamefont{J.}~\bibnamefont{Arenzon}}, \bibnamefont{and}
  \bibinfo{author}{\bibfnamefont{G.}~\bibnamefont{Fabricius}},
  \bibinfo{journal}{Physica A: Statistical Mechanics and its Applications}
  \textbf{\bibinfo{volume}{340}}, \bibinfo{pages}{316} (\bibinfo{year}{2004}).

\bibitem[{\citenamefont{{Denny} et~al.}(2003)\citenamefont{{Denny}, {Reichman},
  and {Bouchaud}}}]{DennyReichman03}
\bibinfo{author}{\bibfnamefont{R.}~\bibnamefont{{Denny}}},
  \bibinfo{author}{\bibfnamefont{D.}~\bibnamefont{{Reichman}}},
  \bibnamefont{and} \bibinfo{author}{\bibfnamefont{J.-P.}
  \bibnamefont{{Bouchaud}}}, \bibinfo{journal}{Phys. Rev. Lett.}
  \textbf{\bibinfo{volume}{90}}, \bibinfo{pages}{025503}
  (\bibinfo{year}{2003}).

\bibitem[{\citenamefont{Angelani et~al.}(1998)\citenamefont{Angelani, Parisi,
  Ruocco, and Viliani}}]{AngelaniParisi98}
\bibinfo{author}{\bibfnamefont{L.}~\bibnamefont{Angelani}},
  \bibinfo{author}{\bibfnamefont{G.}~\bibnamefont{Parisi}},
  \bibinfo{author}{\bibfnamefont{G.}~\bibnamefont{Ruocco}}, \bibnamefont{and}
  \bibinfo{author}{\bibfnamefont{G.}~\bibnamefont{Viliani}},
  \bibinfo{journal}{Phys. Rev. Lett.} \textbf{\bibinfo{volume}{81}},
  \bibinfo{pages}{4648} (\bibinfo{year}{1998}).

\bibitem[{\citenamefont{Doye}(2001)}]{Doye02}
\bibinfo{author}{\bibfnamefont{J.~P.~K.} \bibnamefont{Doye}},
  \bibinfo{journal}{Phys. Rev. Lett.} \textbf{\bibinfo{volume}{88}},
  \bibinfo{pages}{238701} (\bibinfo{year}{2001}).

\bibitem[{\citenamefont{M{\'e}zard et~al.}(1987)\citenamefont{M{\'e}zard,
  Parisi, and Virasoro}}]{MezardParisi87b}
\bibinfo{author}{\bibfnamefont{M.}~\bibnamefont{M{\'e}zard}},
  \bibinfo{author}{\bibfnamefont{G.}~\bibnamefont{Parisi}}, \bibnamefont{and}
  \bibinfo{author}{\bibfnamefont{M.~A.} \bibnamefont{Virasoro}},
  \emph{\bibinfo{title}{Spin-Glass Theory and Beyond}},
  vol.~\bibinfo{volume}{9} of \emph{\bibinfo{series}{Lecture Notes in Physics}}
  (\bibinfo{publisher}{World Scientific}, \bibinfo{address}{Singapore},
  \bibinfo{year}{1987}).

\bibitem[{\citenamefont{Cavagna et~al.}(1999)\citenamefont{Cavagna, Garrahan,
  and Giardina}}]{CavagnaGarrahan99}
\bibinfo{author}{\bibfnamefont{A.}~\bibnamefont{Cavagna}},
  \bibinfo{author}{\bibfnamefont{J.}~\bibnamefont{Garrahan}}, \bibnamefont{and}
  \bibinfo{author}{\bibfnamefont{I.}~\bibnamefont{Giardina}},
  \bibinfo{journal}{Phys. Rev. E} \textbf{\bibinfo{volume}{59}},
  \bibinfo{pages}{2808} (\bibinfo{year}{1999}).

\bibitem[{\citenamefont{Fyodorov}(2004)}]{Fyodorov04}
\bibinfo{author}{\bibfnamefont{Y.}~\bibnamefont{Fyodorov}},
  \bibinfo{journal}{Phys. Rev. Lett.} \textbf{\bibinfo{volume}{92}},
  \bibinfo{pages}{240601} (\bibinfo{year}{2004}).

\bibitem[{\citenamefont{{Burda} et~al.}(2006)\citenamefont{{Burda},
  {Krzywicki}, {Martin}, and {Tabor}}}]{BurdaKrzywicki06}
\bibinfo{author}{\bibfnamefont{Z.}~\bibnamefont{{Burda}}},
  \bibinfo{author}{\bibfnamefont{A.}~\bibnamefont{{Krzywicki}}},
  \bibinfo{author}{\bibfnamefont{O.~C.} \bibnamefont{{Martin}}},
  \bibnamefont{and} \bibinfo{author}{\bibfnamefont{Z.}~\bibnamefont{{Tabor}}},
  \bibinfo{journal}{Phys. Rev. E} \textbf{\bibinfo{volume}{73}},
  \bibinfo{pages}{036110} (\bibinfo{year}{2006}).

\bibitem[{\citenamefont{{De Dominicis} and
  Goldschmidt}(1989)}]{DominicisGoldschmidt89}
\bibinfo{author}{\bibfnamefont{C.}~\bibnamefont{{De Dominicis}}}
  \bibnamefont{and}
  \bibinfo{author}{\bibfnamefont{Y.}~\bibnamefont{Goldschmidt}},
  \bibinfo{journal}{J. Phys. A} \textbf{\bibinfo{volume}{22}},
  \bibinfo{pages}{L775} (\bibinfo{year}{1989}).

\bibitem[{\citenamefont{Wales}(2003)}]{Wales03}
\bibinfo{author}{\bibfnamefont{D.}~\bibnamefont{Wales}},
  \emph{\bibinfo{title}{Energy Landscapes}} (\bibinfo{publisher}{Cambridge
  University Press}, \bibinfo{address}{Cambridge}, \bibinfo{year}{2003}).

\bibitem[{\citenamefont{Doye and Massen}(2005)}]{DoyeMassen05}
\bibinfo{author}{\bibfnamefont{J.~P.~K.} \bibnamefont{Doye}} \bibnamefont{and}
  \bibinfo{author}{\bibfnamefont{C.~P.} \bibnamefont{Massen}},
  \bibinfo{journal}{Phys. Rev. E} \textbf{\bibinfo{volume}{71}},
  \bibinfo{pages}{016128} (\bibinfo{year}{2005}).

\bibitem[{\citenamefont{Li}(1981)}]{Li81}
\bibinfo{author}{\bibfnamefont{T.}~\bibnamefont{Li}}, \bibinfo{journal}{Phys.
  Rev. B} \textbf{\bibinfo{volume}{24}}, \bibinfo{pages}{6579}
  (\bibinfo{year}{1981}).

\bibitem[{\citenamefont{Derrida and Gardner}(1986)}]{DerridaGardner86c}
\bibinfo{author}{\bibfnamefont{B.}~\bibnamefont{Derrida}} \bibnamefont{and}
  \bibinfo{author}{\bibfnamefont{E.}~\bibnamefont{Gardner}},
  \bibinfo{journal}{J. Physique} \textbf{\bibinfo{volume}{47}},
  \bibinfo{pages}{959} (\bibinfo{year}{1986}).

\bibitem[{\citenamefont{Derrida and Weisbuch}(1987)}]{DerridaWeisbuch87}
\bibinfo{author}{\bibfnamefont{B.}~\bibnamefont{Derrida}} \bibnamefont{and}
  \bibinfo{author}{\bibfnamefont{G.}~\bibnamefont{Weisbuch}},
  \bibinfo{journal}{Europhys. Lett.} \textbf{\bibinfo{volume}{4}},
  \bibinfo{pages}{657} (\bibinfo{year}{1987}).

\bibitem[{\citenamefont{Fabricius and Stariolo}(2002)}]{FabriciusStariolo02}
\bibinfo{author}{\bibfnamefont{G.}~\bibnamefont{Fabricius}} \bibnamefont{and}
  \bibinfo{author}{\bibfnamefont{D.~A.} \bibnamefont{Stariolo}},
  \bibinfo{journal}{Phys. Rev. E} \textbf{\bibinfo{volume}{66}},
  \bibinfo{pages}{031501} (\bibinfo{year}{2002}).

\bibitem[{\citenamefont{Bertin}(2005)}]{Bertin05}
\bibinfo{author}{\bibfnamefont{E.}~\bibnamefont{Bertin}},
  \bibinfo{journal}{Europhys. Lett.} \textbf{\bibinfo{volume}{71}},
  \bibinfo{pages}{452} (\bibinfo{year}{2005}).

\bibitem[{\citenamefont{Montanari and Semerjian}(2006)}]{MontanariSemerjian06}
\bibinfo{author}{\bibfnamefont{A.}~\bibnamefont{Montanari}} \bibnamefont{and}
  \bibinfo{author}{\bibfnamefont{G.}~\bibnamefont{Semerjian}},
  \bibinfo{journal}{J. Stat. Phys.} \textbf{\bibinfo{volume}{124}},
  \bibinfo{pages}{103} (\bibinfo{year}{2006}).

\bibitem[{\citenamefont{Barabasi and Albert}(1999)}]{BarabasiAlbert99}
\bibinfo{author}{\bibfnamefont{L.}~\bibnamefont{Barabasi}} \bibnamefont{and}
  \bibinfo{author}{\bibfnamefont{R.}~\bibnamefont{Albert}},
  \bibinfo{journal}{Science} \textbf{\bibinfo{volume}{286}},
  \bibinfo{pages}{509} (\bibinfo{year}{1999}).

\bibitem[{\citenamefont{{Guimaraes, Jr.} et~al.}(2005)\citenamefont{{Guimaraes,
  Jr.}, {de Aguiar}, Bascompte, Jordano, and {Furtado dos
  Reis}}}]{GuimaraesdeAguiar05}
\bibinfo{author}{\bibfnamefont{P.~R.} \bibnamefont{{Guimaraes, Jr.}}},
  \bibinfo{author}{\bibfnamefont{M.~A.} \bibnamefont{{de Aguiar}}},
  \bibinfo{author}{\bibfnamefont{J.}~\bibnamefont{Bascompte}},
  \bibinfo{author}{\bibfnamefont{P.}~\bibnamefont{Jordano}}, \bibnamefont{and}
  \bibinfo{author}{\bibfnamefont{S.}~\bibnamefont{{Furtado dos Reis}}},
  \bibinfo{journal}{Phys. Rev. E} \textbf{\bibinfo{volume}{71}},
  \bibinfo{pages}{037101} (\bibinfo{year}{2005}).

\bibitem[{\citenamefont{Bak et~al.}(1988)\citenamefont{Bak, Tang, and
  Wiesenfeld}}]{BakChao88}
\bibinfo{author}{\bibfnamefont{P.}~\bibnamefont{Bak}},
  \bibinfo{author}{\bibfnamefont{C.}~\bibnamefont{Tang}}, \bibnamefont{and}
  \bibinfo{author}{\bibfnamefont{K.}~\bibnamefont{Wiesenfeld}},
  \bibinfo{journal}{Phys. Rev. A} \textbf{\bibinfo{volume}{38}},
  \bibinfo{pages}{364} (\bibinfo{year}{1988}).

\bibitem[{\citenamefont{Sciortino et~al.}(1999)\citenamefont{Sciortino, Kob,
  and Tartaglia}}]{SciortinoKob99}
\bibinfo{author}{\bibfnamefont{F.}~\bibnamefont{Sciortino}},
  \bibinfo{author}{\bibfnamefont{W.}~\bibnamefont{Kob}}, \bibnamefont{and}
  \bibinfo{author}{\bibfnamefont{P.}~\bibnamefont{Tartaglia}},
  \bibinfo{journal}{Phys. Rev. Lett.} \textbf{\bibinfo{volume}{83}},
  \bibinfo{pages}{3214} (\bibinfo{year}{1999}).

\bibitem[{\citenamefont{Sastry}(2001)}]{Shastry01}
\bibinfo{author}{\bibfnamefont{S.}~\bibnamefont{Sastry}},
  \bibinfo{journal}{Nature} \textbf{\bibinfo{volume}{409}},
  \bibinfo{pages}{164} (\bibinfo{year}{2001}).

\bibitem[{\citenamefont{Stillinger and Weber}(1982)}]{StillingerWeber82}
\bibinfo{author}{\bibfnamefont{F.~H.} \bibnamefont{Stillinger}}
  \bibnamefont{and} \bibinfo{author}{\bibfnamefont{T.~A.} \bibnamefont{Weber}},
  \bibinfo{journal}{Phys. Rev. A} \textbf{\bibinfo{volume}{25}},
  \bibinfo{pages}{978} (\bibinfo{year}{1982}).

\bibitem[{\citenamefont{Weinrib and Halperin}(1982)}]{WeinribHalperin82}
\bibinfo{author}{\bibfnamefont{A.}~\bibnamefont{Weinrib}} \bibnamefont{and}
  \bibinfo{author}{\bibfnamefont{B.}~\bibnamefont{Halperin}},
  \bibinfo{journal}{Phys. Rev. B} \textbf{\bibinfo{volume}{26}},
  \bibinfo{pages}{1362} (\bibinfo{year}{1982}).

\end{thebibliography}

\end{document}